\documentclass[utf8]{FrontiersinHarvard} 

\usepackage{url,hyperref,microtype,subcaption}
\usepackage[onehalfspacing]{setspace}

\newcommand\apj{Astrophys. J.}

\newcommand\aap{Astron. Astrophys.}
\newcommand\solphys{Sol. Phys.}

\def\keyFont{\fontsize{8}{11}\helveticabold }
\def\firstAuthorLast{Wang {et~al.}} 
\def\Authors{Rui Wang\,$^{1,2,*}$, Yiming Jiao\,$^{1,2}$, Xiaowei Zhao\,$^{3}$, and Chong Huang\,$^{4}$}
\def\Address{$^{1}$ State Key Laboratory of Solar Activity and Space Weather, National Space Science Center, Chinese Academy of Sciences, Beijing 100190, People's Republic of China \\
$^{2}$ University of Chinese Academy of Sciences, Beijing 100049, China \\
$^{3}$ Key Laboratory of Space Weather, National Satellite Meteorological Center (National Center for Space Weather), China Meteorological Administration, Beijing 100081, People's Republic of China \\
$^{4}$ School of Electronic Information and Electrical Engineering, Huizhou University, Huizhou 510609, China}
\def\corrAuthor{Corresponding Author}

\def\corrEmail{rwang@swl.ac.cn}

\begin{document}
\onecolumn
\firstpage{1}

\title[Multi-scale Energy Release Events in the Quiet Sun]{Multi-scale Energy Release Events in the Quiet Sun: A Possible Source for Coronal Heating} 

\author[\firstAuthorLast ]{\Authors} 
\address{} 
\correspondance{} 

\extraAuth{}

\maketitle

\begin{abstract}

\section{}
The coronal heating problem remains one of the most challenging questions in solar physics. The energy driving coronal heating is widely understood to be associated with convective motions below the photosphere. Recent high-resolution observations reveal that photospheric magnetic fields in the quiet Sun undergo complex and rapid evolution. These photospheric dynamics are expected to be reflected in the coronal magnetic field. Motivated by these insights, our research aims to explore the relationship between magnetic energy and coronal heating. By combining observations from Solar Orbiter and SDO with a magnetic field extrapolation technique, we estimate the magnetic free energy of multi-scale energy release events in the quiet Sun. Interestingly, our results reveal a strong correlation between the evolution of free energy and the integrated intensity of extreme ultraviolet emission at 171 \AA~in these events. We quantitatively assess the potential energy flux budget of these events to evaluate their contribution to coronal heating. Our study implies a link between photospheric magnetic field evolution and coronal temperature variations, paving the way for further research into similar phenomena.

\tiny
 \keyFont{ \section{Keywords:} Quiet solar corona (1992), Solar coronal heating (1989), Solar coronal transients (312), Solar extreme ultraviolet emission (1493), Solar magnetic reconnection (1504)} 
\end{abstract}

\section{Introduction}

The coronal heating problem remains one of the most fundamental unsolved mysteries in solar physics and space physics in the 21st century \citep{2006Klimchuk,2012Parnell,2014Priest,2015Klimchuk}. A fundamental paradox arises from the remarkable temperature disparity between the solar corona and its underlying photosphere. Namely, the photosphere maintains a temperature of approximately 6000 K, while the outer corona reaches temperatures of 1-2 MK, separated only by a thin transition region of a few hundred kilometers \citep{2009Yangsh,2014Priest,2015Moortel}. This extraordinary temperature gradient challenges fundamental physical principles. It is well established that the energy source originates from convective motions below the photosphere. Then how is the energy transferred through the cooler photosphere to the hotter corona, as the apparent upward heat flow from cooler to hotter regions seems to contradict the Second Law of Thermodynamics. The precise physical process behind this dramatic temperature increase remains elusive. The challenge extends beyond merely explaining the high temperatures. Although the corona's low density (typically 10$^8$-10$^9$ cm$^{-3}$) results in smaller energy flux, which helps maintain its high temperature, the continuous radiative and conductive losses ($\sim$3$\times$10$^5$ erg cm$^{-2}$ s$^{-1}$ in quiet Sun and $\sim$10$^7$ erg cm$^{-2}$ s$^{-1}$ in active regions) to the surrounding cooler atmosphere require a persistent energy supply \citep{1977Withbroe,2004Aschwanden,2014Reale}. Observations from various space-based instruments show that coronal temperatures remain remarkably stable over various temporal and spatial scales \citep{2014Reale,2018Del}, indicating continuous heating is necessary to prevent the rapid cooling. Therefore, solving the coronal heating problem requires not only explaining the extremely high temperatures but also identifying the specific physical mechanisms that sustain this high-temperature state.

In simple terms, the coronal heating mechanisms are generally categorized into two main types: wave-based heating and nanoflare-driven heating. Wave-based heating, primarily through Alfv$\acute{e}$n and magnetoacoustic waves, offers an explanation for continuous heating with efficient energy transport \citep{1978Ionson,1983Heyvaerts}. However, this mechanism faces challenges including chromospheric damping, mode conversion processes, and the identification of precise dissipation mechanisms in the corona such as phase mixing and resonant absorption \citep{2020Van}. The nanoflare-driven heating mechanism, proposed by \citet{1988Parker}, attributes coronal heating to magnetic braiding and subsequent reconnection events triggered by photospheric convective motions. While this model effectively explains impulsive heating and observations of high-temperature plasma \citep{2006Klimchuk}, questions remain regarding the frequency of nanoflares, their spatial-temporal distribution, and whether their total energy budget is sufficient for sustained coronal heating \citet{2018Priest}. Contemporary theoretical frameworks suggest that these mechanisms likely coexist and interact \citep{2015Moortel}, with wave-based heating potentially providing background heating while nanoflares contribute to intermittent strong heating events. Different mechanisms may dominate in different coronal regions, and their coupling effects could be crucial for maintaining the corona's high temperature state \citep{2015Klimchuk}. The relative importance and interaction of these mechanisms remain active areas of research in solar physics.

It is widely accepted that the energy driving coronal or chromospheric heating associated with convective motions below the photosphere. Recent high-resolution observations reveal the quiet-Sun photosphere to be complex and rapidly evolving. Since photospheric magnetic fields extend into the solar corona, these photospheric complexity and dynamics are expected to be reflected in the coronal magnetic field. Meyer et al. (2013) investigated magnetic energy storage and dissipation in the quiet-Sun corona. They concluded that the magnetic free energy stored in the coronal field is sufficiently abundant to explain small-scale phenomena such as X-ray bright points and other impulsive events, providing crucial insights into the underlying mechanisms of solar coronal heating. In this report, we aim to study the characteristics of free energy evolution in multi-scale quiet-Sun eruptions and their relation to coronal heating, by combining extensive temporal coverage observations from the Solar Dynamics Observatory \citep[SDO;][]{2012Pesnell} with high spatial and temporal resolution observations from the Solar Orbiter \citep[SolO;][]{2020Muller}. Furthermore, we quantitatively assess the potential energy flux of these eruptive events to evaluate their significance in coronal heating. In Section 2, we present the data analysis of the imaging results and magnetic field, and in Section 3, we engage in discussions.

\section{Data Analysis and Results}\label{sec2}
For joint observations, we select the SolO observation data from 2022 March 8, when it was positioned almost midway along the Sun-Earth line ($\sim$0.49 au to the Sun). The extreme ultraviolet (EUV) High Resolution Imager (HRI$_{EUV}$) of the Extreme Ultraviolet Imager \citep[EUI;][]{2020Rochus} on SolO provides continuous observations with a time resolution of 3 s for half an hour starting from 00:00 UT on March 8. The angular pixel size is 0$^{\prime\prime}$.492, corresponding to about 175 km on the solar surface, which is about 2.5 times better than the spatial resolution of the Atmospheric Imaging Assembly \citep[AIA;][]{2012Lemen} on board SDO (0$^{\prime\prime}$.6/pixel, $\sim$440 km on the solar surface). The light travel time difference between the Sun and each instrument is about 250 s. We select three representative small-scale flare events (see F1, F2, and F3 of Animation 1) in the quiet Sun near the disk center. These regions exhibit obvious small-scale solar activities. Their sizes range from a few megameters to 20 Mm. Figure 1a shows an eruption of similar spatial scale to the small-scale dimming events catalogued in Table 1 of \citet{2023Rui} using EUI/HRI$_{EUV}$ observations. Combining the animation and Figure 1b, we find that this should be a failed eruption, where the material is ejected outward and then falls back to the solar surface along the black arc shown in Figure 1a. The animation also shows that there are relatively obvious material ejections in the south of the eruption source. These phenomena suggest that: (1) the material and energy from the quiet-Sun eruptions may have difficulty propagating to higher altitudes, and (2) these small-scale eruptions may be composed of or accompanied by even smaller-scale events.

\begin{figure}[h]
    \centering
    \includegraphics[width=0.8\textwidth]{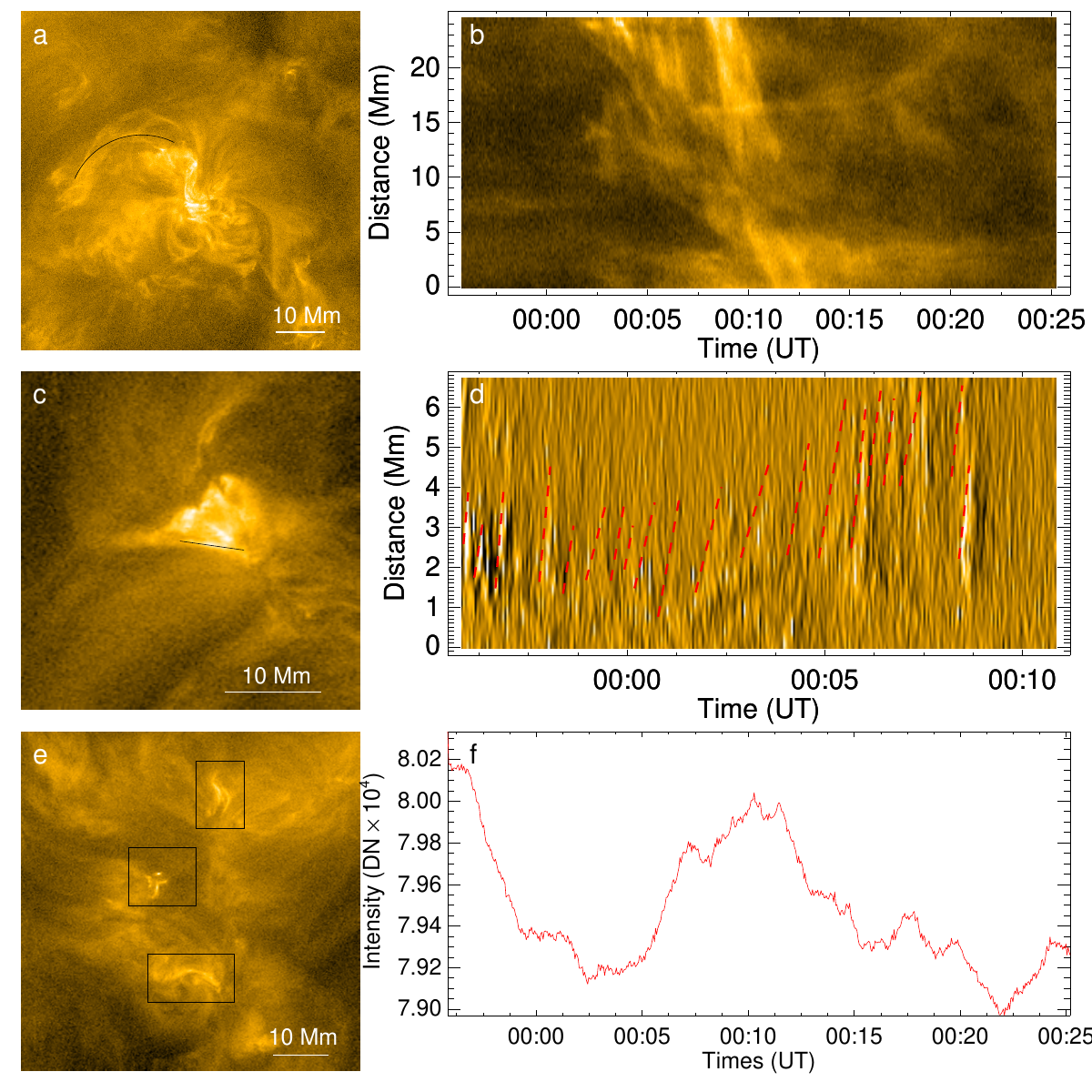}
    \caption{EUI/HRI$_{EUV}$ 174 \AA~observations. (a) A failed small-scale filament eruption. (b) Time-distance plot along the black arc in (a) (right endpoint = 0). (c) A confined eruption with multiple EUV brightenings and mini-jets. (d) Time-distance difference plot along the black line in (c) (right endpoint = 0), with the projected speeds of small jets annotated by red dashed lines. (e) Three smaller-scale flares. (f) Light curve showing the sum of integrated intensities within the three black boxes in (e). All times are in SDO reference frame. F1, F2, and F3 in the animation align with areas shown in panels (a), (c), and (e).}\label{fig1}
\end{figure}

Figure 1c shows a smaller region compared to Figure 1a. F2 does not exhibit obvious filament or flare eruptions, but it does show clear EUV brightenings. Careful examination of the animation reveals jet-like structures along the black line in Figure 1c and its adjacent regions. We believe these jet-like structures are likely associated with outflows related to magnetic reconnection. Figure 1d shows that these jet-like structures have relatively high speeds, and it would have been difficult to identify them without the high temporal resolution of HRI$_{EUV}$. The average speed of 20 small jets within 15 minutes is 113 km s$^{-1}$, with a median of 106 km s$^{-1}$, ranging from 55 km s$^{-1}$ to 200 km s$^{-1}$, which is the typical speed range associated with small-scale jets related to reconnection events \citep{2021Chitta,2023ChengX}. Note that the measured speeds are projected in the plane of the sky. Figure 1e shows multi-scale eruptions, some of which exhibit jet-like features, while others only show EUV brightenings (see F3 of Animation 1). We aim to investigate whether these small-scale solar activities shown in Figure 1 could potentially serve as an energy source for coronal heating.

The energy driving coronal and chromospheric heating is widely understood to be associated with convective motions below the photosphere. The magnetic free energy in the quiet-Sun coronal field, influenced by photospheric magnetic field evolution, reflects energy storage and dissipation dynamics in the quiet-Sun corona. Solar flares and similar energy releases are fundamentally facilitated by magnetic reconnection, which generates observable EUV  brightenings. Motivated by these insights, our research aims to quantify the relationship between temporal variations in emission intensity within eruption source regions and concurrent changes in magnetic free energy, seeking to unravel the intricate mechanisms underlying coronal heating. The AIA offers observations in six channels, with a broader temporal coverage than the HRI$_{EUV}$ data, which facilitates the study of energy evolution. We use data from one hour before and after the period of interest. However, the time resolution of some AIA channels is only 96 s during the period from 23:00 UT on March 7 to 02:00 UT on March 8. Therefore, to align the data across all channels, we uniformly adopte a time interval of 96 s.

\begin{figure}[h]
    \centering
    \includegraphics[width=0.8\textwidth]{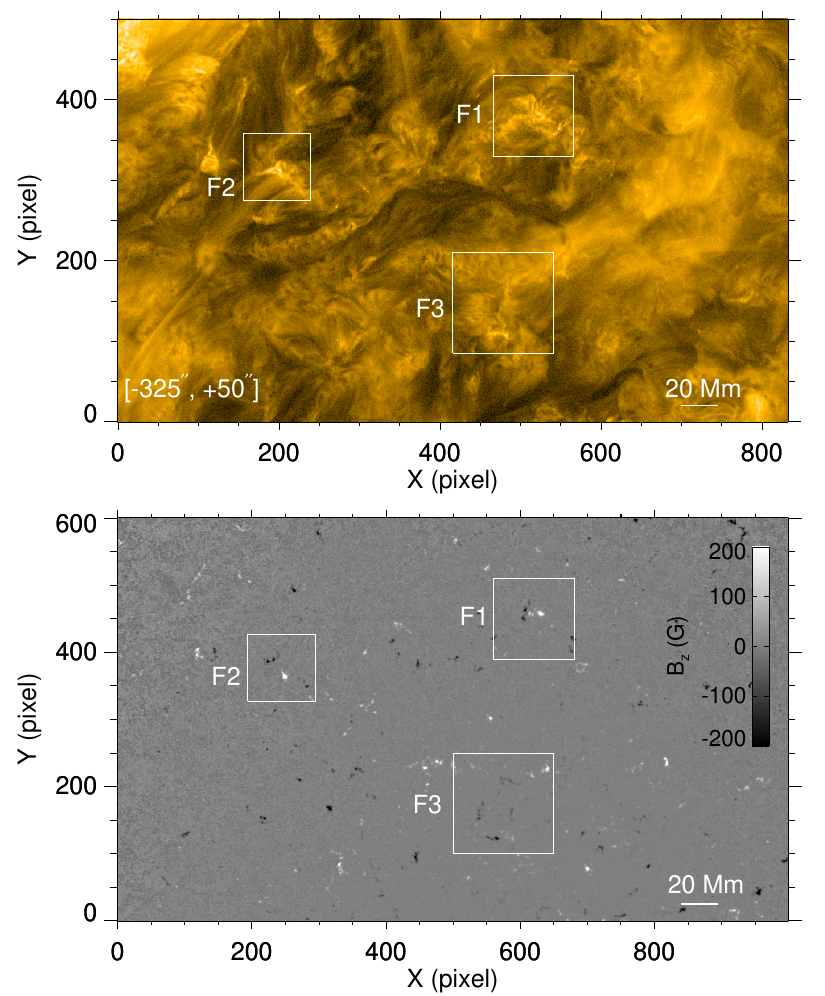}
    \caption{Distribution of small-scale eruptive events. Top: AIA 171 \AA~image showing events F1, F2, and F3 (corresponding to microflares in Figure 1a, 1c, and 1e). White boxes mark the regions for intensity integration in Figure 3. Helioprojective coordinates are shown at the lower-left corner. Bottom: White boxes correspond to those in the top panel overlaid on magnetogram (Bz), which define the free energy calculation regions. Actual extrapolation boundary extends roughly twice the box size.}\label{fig2}
\end{figure}

Figure 2 shows the EUV image at 171 \AA~and photospheric magnetogram in the quiet Sun. The white boxes outline the regions selected for intensity integration and magnetic free energy. To calculate the magnetic free energy, we use a nonlinear force-free (NLFFF) magnetic field extrapolation method \citep{2004Wiegelmann,2012Wiegelmann} to obtain the coronal magnetic field. The Helioseismic and Magnetic Imager \citep[HMI;][]{2012Scherrer,2012Schou} on board SDO does not provide the direct boundary data for NLFFF extrapolation in the quiet Sun. We adopt the ``bvec2cea.pro'' routine in SSW packages to convert the disambiguated full-disk vector magnetic field ``hmi.B\_720s'' series from the native CCD coordinates to the cylindrical equal area heliographic coordinates \citep{2014Hoeksema}, which is appropriate for extrapolation. The ``bvec2cea.pro'' routine uses a radial-acute method \citep{2014Hoeksema} to resolve the 180$^\circ$ azimuthal uncertainty in the transverse field direction. This method has been demonstrated effective and reliable by previous studies \citep{2023Rui}. They successfully reconstruct the magnetic flux-rope structures in the quiet Sun, which show good agreement with observations of the filaments.

Figure 3 demonstrates a strong correlation between the integrated intensity in the 171 \AA~channel (red curve in the second row) and magnetic free energy [$\int \frac{B^2}{2\mu_0}~dV$] (black curve in the third row), with correlation coefficients ranging from 0.60 to approximately 0.86 across the three regions. The integrated intensity curves for all EUV channels are plotted in the first row to detect potential microflares. Similar to active region flares, we detect distinct impulsive intensity enhancements in high-temperature channels (e.g., 131 \AA~ and 94 \AA) in each region. These impulsive peaks largely coincide with the peaks in free energy, suggesting a correlation between energy release and EUV emission intensity. Among all channels, the 171 \AA~intensity curve shows the strongest correlation with free energy variations. Furthermore, we integrate the intensity from higher-resolution HRI$_{EUV}$ images at 174 \AA~and compare it with the light curve of AIA 171 \AA, which displays nearly identical curve patterns (shown in Figure 1f and the dashed box in Figure 3f). This indicates that multiple small-scale eruptions, when integrated, can characterize the intensity variations across the entire region.

\begin{figure}[h]
    \centering
    \includegraphics[width=1.0\textwidth]{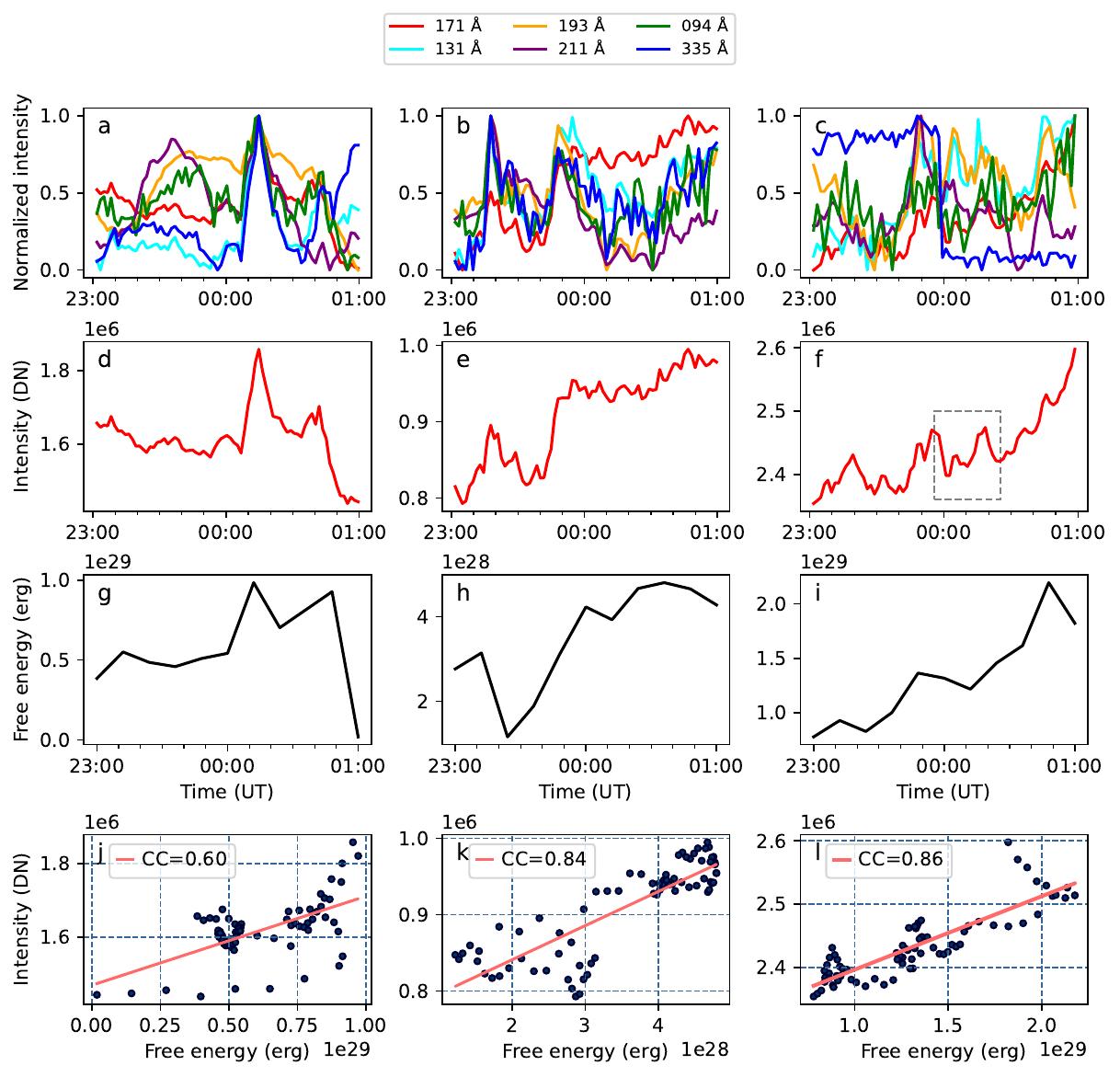}
    \caption{Time evolution of AIA EUV intensity and magnetic free energy. (a)-(c): Normalized intensity curves for six EUV channels. (d)-(f): Intensity curves at 171 \AA. The dashed box in (f) marking the time period shown in Figure 1f. (g)-(i): Magnetic free energy evolution. (j)-(l) Scatter plots of the correlation coefficient between the free energy and the corresponding EUV 171 \AA~intensity above. The linear fit is shown by the red solid line. Columns from left to right correspond to F1, F2, and F3 shown in Figure 2. }\label{fig3}
\end{figure}

\begin{figure}[h]
    \centering
    \includegraphics[width=0.8\textwidth]{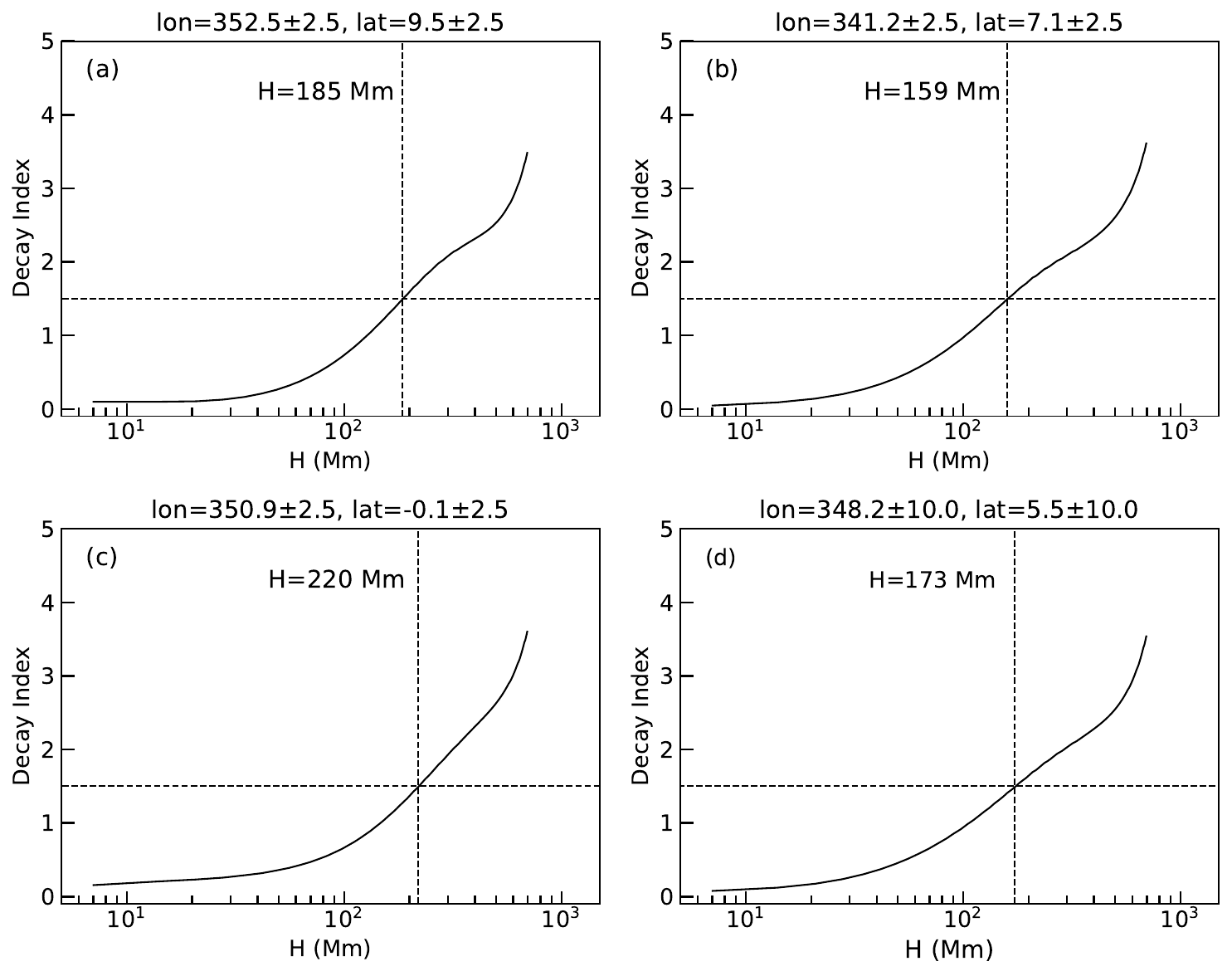}
    \caption{Decay index $n$ from PFSS model. (a-c) Decay index profiles above regions F1, F2, and F3, with calculation ranges in Carrington coordinates shown above each panel. Critical height H for torus instability onset ($n$ = 1.5) is marked. Panel (d) shows the average decay index profile over a larger region encompassing all three events.}\label{fig4}
\end{figure}

We choose the range of EUV intensity to encompass the source region of the eruption as much as possible, such as the box corresponding to F1. In addition to capturing the sigmoid hot channel structures of the eruption core region, we also need to capture the region where the material falls along the arc trajectory as shown in Figure 1a. For the selection of the region F2, besides enclosing the wizard hat structure, we also try to avoid including another similar eruption structure in the west, so the region cannot be chosen too large. For the region F3, we choose the three neighboring eruption regions. Figure 1f reveals that the overall intensity variation is mainly determined by these three major eruptions occurring at different times. As for the calculation of the magnetic energy, we keep the calculation regions of magnetic energy and EUV intensity as consistent as possible, which encompass the main magnetic poles related to the eruptions.

In fact, the change in region size does not significantly affect the trend of magnetic field or EUV intensity changes, but only their amplitudes. Through our investigations, we find that as long as the core region of the eruption source is included, the overall trend of intensity and magnetic energy changes will not change too much, and the larger the region, the smaller the impact on the amplitude when scaling the region area, because most of the contribution comes from the intensity and magnetic energy of the core region.

Compared to the 171 \AA~channel, the correlation with other channels is not as strong. We attribute this primarily to the characteristic emission temperatures of different types of solar features. The 171 \AA~exhibits a better temperature response to the quiet corona, while the 131 \AA~and 94 \AA~are more sensitive to the flaring corona \citep{2012Lemen}. We observe that F1, a small-scale filament eruption with flaring characteristics located in a quiet region, contrasts with F2, which shows no sigificant eruption but rather a slow release of energy through tiny jets driven by reconnection. The magnetic energy of F2 may accumulate through quasi-static processes. AIA 171 \AA~observations are more effective in capturing such features during non-eruptive periods, whereas the 131 \AA~and 94 \AA~channels are better suited for detecting high-temperature plasma during impulsive flares. As a result, F2 shows a stronger correlation between magnetic energy and EUV intensity at 171 \AA~compared to the 131 \AA~and 94 \AA~channels, and this difference in correlation is more pronounced than in region F1.

\section{Discussion} 
Our analysis reveals an interesting finding: the integrated EUV intensity shows strong correlation with magnetic free energy within the specified time intervals. The three regions exhibit eruptions of different scales. F1 corresponds to a failed small-scale filament eruption, where we believe the magnetic free energy is partially converted into thermal energy through dissipation, while a significant portion transform into kinetic energy of the upward-moving material, which eventually falls back to the solar surface due to gravitational and magnetic confining force. Figure 4 illustrates the magnetic confining force changes with height above each eruption source region, characterized by the decay index \citep{2006Kliem}. We use PFSS model to calculate the decay index \citep{2020Stansby}. The critical heights for torus instability in all three regions exceed 150 Mm. Such heights are typically beyond the reach of small-scale filaments, which makes torus instability-driven eruptions and subsequent interplanetary mass and energy transport unlikely. In other words, even if the eruption energy is fully converted to kinetic energy, it ultimately dissipates as thermal energy within the corona.

Unlike active region flares, where EUV peaks are followed by significant energy drops, the EUV intensity here reveals an unexpected correlation with magnetic free energy. We propose this occurs because, in contrast to large-scale eruptive events, there is no significant mass or energy transfer outside the computation domain. Moreover, these small-scale eruptions may not significantly impact the photospheric magnetic field causing irreversible changes, as observed in the studies of active region eruptions \citep{1992Wanghm,2000Hudson,2014Rui}. Additionally, the NLFFF extrapolation method cannot adequately capture the nonlinear processes of magnetic reconnection, which prevents the free energy from showing a significant decrease associated with energy release. Nevertheless, we believe that using the continuous increase in free energy to represent magnetic energy deposition is reasonable. In essence, the extrapolated free energy changes primarily reflect the evolution of the photospheric magnetic field, which transfers and deposits energy into coronal magnetic fields. However, this deposited free energy is expected to be actually released through magnetic reconnection to heat the corona. In contrast, EUV integrated intensity depends on real-time emission changes, which typically intensify during flares or magnetic reconnection events through the photospheric motions. The 171 \AA~emission is particularly sensitive to coronal temperature changes, which corresponds to the background magnetic structure of quiet regions. All these above lead to our hypothesis, i.e., the continuous shuffling and intermixing of field footpoints in the photospheric convection causes coronal magnetic fields to wind and interweave. This process continuously leads to energy dissipation through magnetic reconnection of the braiding coronal magnetic fields, with the dissipated energy manifesting as a deposition of the free energy.

This aligns well with the nanoflare hypothesis proposed by \citet{1988Parker}. Therefore, we can expect that continuous smaller-scale magnetic reconnection events, similar in size to the observed quiet-Sun coronal brightenings, may provide a relatively stable, continuous heating source. For instance, the integrated intensity curve of F3 is the result of three smaller-scale eruptive features. The sum of their individual integrated intensity curve (Figure 1f) has almost the same shape with the integrated intensity curve obtained over the entire region (Figure 3f). The intensity curve correlates well with the changes in magnetic free energy, with a correlation coefficient of 0.86. Based on the strong correlation between the free energy and the EUV intensity, we hypothesize that most of the deposited free energy is actually released through magnetic reconnection, which enables us to estimate the energy for heating the corona within these areas by calculating ($\Delta$ E$_{free}$).

Considering only the change from the minimum to the maximum free energy during a time interval, and ignoring the fluctuations within the time interval, we can roughly estimate that the three events could provide energy fluxes of $\sim$1.8$\times$10$^6$ erg cm$^{-2}$ s$^{-1}$, $\sim$7$\times$10$^5$ erg cm$^{-2}$ s$^{-1}$, and $\sim$1.2$\times$10$^6$ erg cm$^{-2}$ s$^{-1}$, respectively. While these values are larger than the average quiet-Sun coronal energy loss rate ($\sim$3$\times$10$^5$ erg cm$^{-2}$ s$^{-1}$; \citealt{1977Withbroe,2006Klimchuk}), we have not excluded the energy from the chromosphere, the loss rate of which can be significantly higher ($\sim$4$\times$10$^6$ erg cm$^{-2}$ s$^{-1}$). Previous similar implementations by \citet{2013Wiegelmann} and \citet{2014Chitta} suggested that the energy was too small to explain coronal heating in the quiet Sun. However, \citet{2013Wiegelmann} used a potential field extrapolation model, which did not take into account the magnetic free energy associated with field tangling and twisting. \citet{2014Chitta} only calculated the energy conducted and radiated below the base of the corona, without considering the energy released through magnetic reconnection in the corona.

On the other hand, our calculations of the energy flux are still relatively crude and could be optimized in several ways, as we do not account for the energy rises and falls within our calculation intervals, which may indicate repeated energy replenishment and release. The relatively coarse temporal resolution (12 minutes) of HMI vector magnetograms could impact the $\Delta$ E$_{free}$ measurements. Using quiet-Sun magnetic field for extrapolation may underestimate free energy estimation. A recent study by \citet{2025Beck} revealed an intriguing finding. By comparing magnetic field measurements from HMI with higher-resolution Hinode SP data, they found that HMI magnetic field data may be significantly underestimated. Specifically, the magnetic field strength in quiet regions with B $<$ 220 G might be underestimated by a factor of 3--10. Furthermore, they suggested that the free energy derived from magnetic field extrapolations across the entire field of view, including both active and quiet regions, could be underestimated by a factor of 2. Given these findings, our estimated free energy of the magnetic field is likely to be substantially lower than the true value. However, considering computational challenges and boundary condition selection that may introduce additional uncertainties, we do not intend to perform the mentioned field correction but may explore it in future research. Nevertheless, the strong correlations between magnetic free energy and EUV 171 \AA~integrated intensity imply a link between photospheric magnetic evolution and coronal temperature changes. We hope to uncover deeper insights into this relationship by examining more similar events in future work.

\section*{Acknowledgments}
We thank the anonymous reviewers for their constructive feedback, which significantly improved the quality of this manuscript. The research was supported by National Key R\&D Program of China (No. 2022YFF0503800), the Strategic Priority Research Program of the Chinese Academy of Sciences (No. XDB0560000), National Natural Science Foundation of china (NSFC, Grant No. 12073032), National Key R\&D Program of China (No. 2021YFA0718600), and the Specialized Research Fund for State Key Laboratories of China. X.W.Z. also acknowledges support from NSFC under grants 42204176, C.H. acknowledges support from the Scientific Research Foundation for the PhD (Huizhou University, 2023JB008). We acknowledge the use of data from Solar Orbiter and SDO. Solar Orbiter is a space mission of international collaboration between ESA and NASA, operated by ESA. The EUI instrument was built by CSL, IAS, MPS, MSSL/UCL, PMOD/WRC, ROB, LCF/IO with funding from the Belgian Federal Science Policy Office (BELSPO/PRODEX PEA 4000134088); the Centre National d'Etudes Spatiales (CNES); the UK Space Agency (UKSA); the Bundesministerium f\"{u}r Wirtschaft und Energie (BMWi) through the Deutsches Zentrum f\"{u}r Luft- und Raumfahrt (DLR); and the Swiss Space Office (SSO).

\bibliographystyle{Frontiers-Harvard} 



\end{document}